\documentclass[twocolumn,showpacs,pra,nofootinbib]{revtex4-2}
\usepackage[T1]{fontenc}
\usepackage[russian,english]{babel}
\usepackage{graphicx}
\usepackage{amsmath}
\usepackage{amssymb}
\usepackage{mathtools}
\usepackage{physics}
\usepackage{color,soul}
\usepackage[colorlinks=true,allcolors=blue]{hyperref}
\usepackage{cancel}
\usepackage{upgreek}
\usepackage[dvipsnames]{xcolor}
\usepackage{slashed}

\graphicspath{{./Figs/}}

\begin{document}

\title{Short-term evolution of electron wave packet in a constant crossed field with radiative corrections}

\date{\today}
\author{I.~Yu.~Kostyukov}
\email[Corresponding author: ]{kost@ipfran.ru}
\author{E.~N.~Nerush}
\affiliation{Institute of Applied Physics of the Russian Academy of Sciences, 46 Ulyanov St., Nizhny Novgorod 603950, Russia}
\author{A.~A.~Mironov} 
\affiliation{LULI, Sorbonne Universit\'e, CNRS, CEA, \'Ecole Polytechnique, Institut Polytechnique de Paris, F-75252 Paris, France}
\affiliation{Theoretical Department, Prokhorov General Physics Institute of the Russian Academy of Sciences, Moscow, 119991, Russia}
\author{A.~M.~Fedotov}
\affiliation{Institute for Laser and Plasma Technologies, National Research Nuclear University MEPhI, 115409, Moscow, Russia}

\begin{abstract}
    We study the dynamics of an electron wave packet in a strong constant crossed electromagnetic field with account for radiative corrections due to interaction of the electron with the vacuum fluctuations. We evaluate a wave packet composed of the solutions to the Dyson-Schwinger equation, which describes electron propagation without emission of real photons. Spacetime dependence of the wave packet is obtained analytically for a short time interval, the more restricted from above the wider is the packet in momentum space. The radiative corrections alter the electron wavefunction, resulting in particular in a damping of the wave packet. The expectation value of the Dirac spin operator also gets modified.
\end{abstract}

\maketitle

\section{Introduction}
\label{sec.Intro}

Advances in laser and accelerator technologies made it possible to study various phenomena in a strong electromagnetic (EM) field under the laboratory conditions. In this context a special attention attract the fundamental processes of strong field quantum electrodynamics (SF QED)~\cite{di2012extremely, fedotov2022high, gonoskov2022charged}. Nonlinear Compton scattering and positron production via multiphoton light-by-light scattering have been observed in the milestone SLAC E-144 experiment where a multi-GeV electron beam was collided with terawatt laser pulses~\cite{bula1996observation,burke1997positron}. More recent experiments~\cite{poder2018experimental,cole2018experimental} have explored the quantum nature of the radiation reaction for the electrons in intense laser field. SF QED effects have been also tested for ultrarelativistic electron beams passing through the crystalline fields~\cite{uggerhoj2005interaction,nielsen2020radiation, nielsen2021experimental}. In order to reach a regime of stronger fields, several projects (ELI, LUXE, FACET, XCELS, SEL), aiming in particular at studying the SF QED effects in beam-beam and laser-beam interactions, have been launched~\cite{heinzl2009exploring,abramowicz2021conceptual,yakimenko2019prospect,shaykin2014prospects,shao2020broad}. 

At the moment QED is considered as the most precise physical theory because its coupling (the fine structure constant) $\alpha = e^2/\hbar c \simeq 1/137$ is rather small, where $e<0$ is the electron electric charge, $\hbar$ is the Planck constant and $c$ is the speed of light. Due to that, the perturbation methods became an efficient tool to calculate the desired quantities with high precision~\cite{berestetskii1982quantum}. This, however, might be not always the case once an extremely strong EM field is involved. There might be a large sector of SF QED where the processes cannot be described by the existing theoretical framework~\cite{fedotov2017conjecture,narozhny1980expansion,ritus1972radiative}. Such sector corresponds to a fully nonperturbative regime of SF QED for which the radiative corrections become significant and have to be resummed rather than considered as perturbations. This feature is specific for the strong-field limit \cite{podszus2019high, ilderton2019note}. Whilst some progress was made for configurations with a constant magnetic field \cite{loskutov1981behavior, gusynin1995dimensional, gusynin1999electron} and a constant crossed field \cite{mironov2020resummation, mironov2022structure}, the full resummation is yet to be accomplished. Moreover, calculations of the transition amplitudes for most of high-order processes even when the perturbation methods remain applicable are far from completion \cite{fedotov2022high, gonoskov2022charged}.

In general, a wavefunction (or a density matrix if the mixed states are of interest) provides a complete description of a quantum system. The number of known field configurations allowing explicit solutions is limited and include the Coulomb field \cite{berestetskii1982quantum}, constant homogeneous electric and magnetic fields \cite{bagrov2014dirac}, plane wave and related configurations \cite{ritus1985quantum, di2021wkb}. A special case of the latter is a constant crossed homogeneous EM field. It naturally arises in a locally constant crossed field approximation (LCFA) \cite{fedotov2022high} as describing interactions of ultra-relativistic particles in a general  extremely strong EM field. In this work, we focus on such a field implying the context of the LCFA. 

The evolution of a one-electron (or one-positron) wavefunction $\psi$ in the field of a plane wave is given by the Dirac equation \cite{berestetskii1982quantum}
\begin{equation}\label{eq:Dirac}
		\left[\hat{\slashed{p}}-e \slashed{A} \left(\varphi\right)-m\right]\psi  =  0,
\end{equation}
where $x^\mu = (t,\mathbf{r}) = (t,x,y,z)$, $\hat{p}^\mu = i \partial^{\mu} $ is the 4-momentum operator, $A^\mu (\varphi)$ is the  4-potential of the external EM field, $\varphi=k\cdot x= \omega t-\mathbf{k r}$ is the phase of the particle in the field,   $k^\mu=(\omega,\mathbf{k})$ is the wave 4-vector such that $k^2 = k\cdot A=0 $, and $m$ is the electron mass, ``slash'' indicates the contraction, e.g. $\slashed{A} = \gamma^\mu A_\mu $, with the Dirac gamma matrices $\gamma^\mu$. Hereinafter, we use the natural units $\hbar = c = 1$.

The exact solutions to equation \eqref{eq:Dirac} are known as the Volkov functions~\cite{wolkow1935klasse}. They can be represented as follows
\begin{eqnarray}
	\label{Volkov_state}
	\psi^{(\pm)}_{\mathbf{p}\sigma} & = & \left( 1 \pm\frac{e\slashed{k}\slashed{A}}{2k\cdot p} \right)u_{\mathbf{p}\sigma}^{(\pm)} \exp\left(i \Phi^{e,p}_V \right), \\
	\Phi^{(\pm)}_V  & = &  \mp  p\cdot x -  \intop_{-\infty}^{\varphi} \left(\frac{eA\cdot p}{k\cdot p} \mp\frac{e^{2}A^{2}}{2k\cdot p}\right)d\varphi,
	\label{fv}
\end{eqnarray}
where the upper and lower signs correspond to the solutions with positive and negative energy, respectfully. The 4-spinors $u^{(\pm)}_{\mathbf{p}\sigma }$ are the same as for the field-free Dirac equation. They are characterized by the generalized 4-momentum $p^\mu = (\varepsilon, \mathbf{p})$, where on the mass shell $\varepsilon= \sqrt{\mathbf{p}^2+m^2}$. The spin index $\sigma = \pm1$ corresponds to the states with the spin projections $\pm1/2$ onto the direction along a certain unit vector $\mathbf{n}_0$ in a particle rest frame. The corresponding unit 4-pseudovector $n^\mu = \left(0,\;\mathbf{n}_0\right)$ ($n^2 = -1$) in the laboratory reference frame takes the  form $n^\mu = \left(\mathbf{p} \mathbf{n}_0 / m, \; \mathbf{n}_0 + \mathbf{p} (\mathbf{n}_0 \mathbf{p} ) / m (\varepsilon + m) \right)$, such that $n \cdot p = 0$.   

The field-free spinor $u^{(\pm)}_{\mathbf{p}\sigma }$ can be generated from an arbitrary 4-spinor $w$ by applying the (non-normalized) projection operators $\mathcal{D}_{0}^{(\pm)}= m\pm\slashed{p}$ and $\mathcal{L}_{\sigma}(n)$~\cite{akhiezer1953quantum} 
\begin{eqnarray}
u^{(\pm)}_{\mathbf{p}\sigma} &=& \mathcal{D}_{0}^{(\pm)}\mathcal{L}_{\sigma} (n) w,  
\label{soldir}
\\
\mathcal{L}_{\sigma} (n) &=& 1 +\sigma \gamma^{5}\slashed{n}, 
\label{Ln}
\end{eqnarray}
where $ \gamma^{5}\slashed{n}$ with $\gamma^5 = i \gamma^0 \gamma^1 \gamma^2 \gamma^3 $ is the spin projection operator, such that $ \gamma^{5}\slashed{n} u^{(\pm)}_{\mathbf{p}\sigma} = \sigma u^{(\pm)}_{\mathbf{p}\sigma}$~\cite{ritus1985quantum}.
The spinors $u^{(\pm)}_{\mathbf{p}\sigma}$ are normalized so that $\overline{u}_{\mathbf{p}\sigma}^{(\pm)} u_{\mathbf{p}\sigma}^{(\pm)} = \pm 1$ and $\overline{u}_{\mathbf{p}\sigma}^{(\pm)} \gamma^\mu u_{\mathbf{p}\sigma}^{(\pm)} = p^\mu/m$, where overline stands for the Dirac adjoint of a Dirac spinor $\overline{u} =  u^\dagger \gamma^0$.

The Volkov functions are widely used to describe the strong-field phenomena: field ionization~\cite{reiss1990complete}, high harmonic generation~\cite{becker1997unified,krausz2009attosecond}, and strong-field QED processes~\cite{di2012extremely}. However, in real cases the wavefunction is localized and has a form of a wave packet. The calculation of its evolution by numerical integration of the Dirac equation is  challenging~\cite{mocken2004quantum,mocken2008fft,hammer2014dispersion}. However, the wave packet in a plane-wave field can be represented as a superposition of the Volkov states~\cite{ritus1985quantum}
  \begin{equation}
 	\varPsi\left(\mathbf{r},t\right) =  \int d\mathbf{p} \sum_{\sigma} \left[c_{\mathbf{p}\sigma}^{(+)} \psi_{\mathbf{p}\sigma}^{(+)}\left(\mathbf{r},t\right)+c_{\mathbf{p}\sigma}^{(-)} \psi_{\mathbf{p}\sigma}^{(-)}\left(\mathbf{r},t\right)\right].
 \end{equation}
Their analytical representation \eqref{Volkov_state} strongly simplifies the computation of the packet dynamics~\cite{san2000evolution}.
The amplitudes $c_{\mathbf{p}\sigma}^{(\pm)}$ in this superposition set the wave packet distribution in the momentum space.
Two approaches have been exploited in the recent works. In the first one the amplitudes were obtained by projecting the initial wavefunction $\varPsi\left(\mathbf{r},t_{0}\right)$ onto the Volkov functions at $t=t_0$~\cite{san2000evolution,san2001relativistic,ooi2020squeezed} 
 \begin{equation}
 	c^{(\pm)}_{\mathbf{p}\sigma} =  \int d \mathbf{r} {\psi_{\mathbf{p}\sigma}^{(\pm)}}^\dagger\left(\mathbf{r},t_0\right)\varPsi\left(\mathbf{r},t_{0}\right).
 \end{equation}
In the second approach the amplitudes $c_{\mathbf{p}\sigma}^{(\pm)}$ are specified ad hoc~\cite{peatross2007electron}.  

In the limit of extremely strong EM field the radiative corrections describing self-interaction have to be taken into account~\cite{ritus1972radiative,di2012extremely,fedotov2017conjecture}. The wavefunction with radiative corrections obeys the Dyson-Schwinger equation, which differs from the Dirac equation by an additional term containing the electron mass operator~\cite{berestetskii1982quantum,ritus1985quantum,ritus1972radiative}. A generalization of the Volkov states to account for radiative corrections have been studied for the constant crossed field~\cite{ritus1972radiative} and plane-wave configuration~\cite{meuren2011quantum,podszus2021first}. The analysis reveals radiative damping (or ``decay'') of the electron states  \cite{ritus1972radiative,tamburini2021efficient,podszus2021first,podszus2022nonlinear}. Actually, the Dyson-Schwinger equation applies only to the radiativeless part of the one-electron state, but a part of the initial wavefunction eventually turns into a multi-particle sector combining the electron and the emitted photons. This results in damping of the probability for the electron to stay in a radiativeless state. 

The wave packet dynamics with account for radiative corrections has not been discussed yet. Here we explore this fundamental problem for a constant crossed EM field configuration. In principle, evolution of a wave packet could be evaluated numerically \cite{braun1999numerical,mocken2008fft,fillion2014split,beerwerth2015krylov}. However, numerical integration of the time-dependent multidimensional Dirac equation  still remains challenging \cite{fillion2012numerical,almquist2014high,di2012extremely}. Inclusion of radiative corrections makes it even harder. An analytical approach complements the numerical calculations and is useful to verify them. For a crossed field configuration it is possible to derive an approximate closed-form analytical expression for the wavefunction and to discuss the impact of radiative corrections on the wave packet dynamics. We confine ourselves to a Gaussian wave packet, for which the evolution is further simplified. This still requires multidimensional integration of rapidly oscillating functions, hence some additional approximations. 

The paper is organized as follows. In Sec.~\ref{sec2} we discuss  following Ref.~\cite{ritus1972radiative} the generalized Volkov states with account for radiative corrections for a crossed field configuration. In Sec.~\ref{sec3} we construct a Gaussian wave packet using its decomposition into the generalized Volkov functions. The expectation value of the Dirac spin operator is calculated in Sec.~\ref{sec4}. The results are discussed in Sec.~\ref{sec5}. The proof of commutativity of the spin and positive energies projectors is relegated to Appendix~\ref{appendix}.

\section{Electron states with account for radiative corrections}
\label{sec2}

Let us start with a discussion of the generalized Volkov states in a constant crossed field with account for radiative corrections following Ref.~\cite{ritus1972radiative}. With account for radiative corrections the one-particle wave function obeys the Dyson-Schwinger (sometimes also called Dirac-Schwinger) equation~\cite{dyson1949s,schwinger1951green,fradkin1956quantum,ritus1972radiative}
\begin{equation}
	\left[\hat{\slashed{p}}-e\slashed{A}\left(\varphi\right)-m\right]\psi(x)  =  \int M\left(x,x'\right)\psi(x')d^{4}x',
	\label{eq_x}
\end{equation}
where $M\left(x,x'\right)$ is the renormalized mass-operator in an external EM field. In a lack of its exact explicit expression at certain point we have confine to a one-loop approximation studied in Ref.~\cite{ritus1972radiative}. 
The 4-potential of the external constant crossed field $\mathbf{E} = \mathbf{e}_{x} \xi E_{cr} $ and $\mathbf{B} = \mathbf{e}_{y} \xi E_{cr} $ can be chosen as follows 
\begin{equation}
A_{\mu}  = \frac{\xi E_{cr} \varphi}{m} \left(0,1,0,0\right),
\end{equation}
where the wave vector is set as $k^{\mu}  =  m \left(1,0,0,1 \right)$, so that $\varphi/m =   t-z  $. Here, $\xi$ denotes the field amplitude in units of the QED critical field $E_{cr} = m^2 / |e| \simeq 1.16\times 10^{16}  V/$cm~\cite{di2012extremely,berestetskii1982quantum,ritus1985quantum}.

The Volkov states \eqref{Volkov_state} can be abbreviated as $\psi^{(\pm)}_{\mathbf{p}\sigma} = E_{\pm p} (x) u^{(\pm)}_{\mathbf{p} \sigma}$. Explicitly, the $E_p$-matrix reads:
\begin{eqnarray}
	E_{p} & = & \left[1 -  \Lambda \frac{ \varepsilon \varphi}{2 p_-} \left(\gamma^0 - \gamma^3 \right) \gamma^1 \right] \exp\left(i \Phi_V \right),
	\label{ep-cross}\\
	\Phi_V & = & \mathbf{p r} - p^0 t + \Lambda \frac{ \varepsilon  p_x \varphi^2}{2 m p_- } - \Lambda^2 \frac{ \varepsilon^2 \varphi^3 }{6 m p_-},
	\label{phase}
\end{eqnarray}
where $\Lambda = \xi m /   \varepsilon  $, and $p_- = k \cdot p /m  = p^0 - p_z$. Note that $E_{-p}=\overline{E}_{p}=\gamma^0E^\dagger_p \gamma^0$.

The spin projection operator for the Volkov states has the same form as in Eq.~\eqref{Ln} if the polarization 4-pseudovector $n_\mu$ is replaced by $n_V^\mu$ defined as follows~\cite{ritus1972radiative,ritus1985quantum}
\begin{eqnarray}
n_V^\mu &=& n^\mu - e A^\mu \frac{ \left( k \cdot n \right)}{ k \cdot p}
\nonumber
 \\
&+& k^\mu \left[ \frac{e  \left( A \cdot n \right)}{ k \cdot p} - \frac{e^2 A^2 \left( k \cdot n \right) }{2 \left(k \cdot p \right)^2 } \right], 
\end{eqnarray}
so that $ \gamma^{5}\slashed{n}_V E_p u_{\mathbf{p} \sigma}^{(\pm)} = \sigma E_p u_{\mathbf{p} \sigma}^{(\pm)}$.

The following identities for the $E_p$-matrices (valid also for an off-shell 4-momentum) 
\begin{eqnarray}
\left(\hat{\slashed{p}}-e\slashed{A} \right)  E_{p}(x)  &=&  E_{p}(x) \slashed{p}, \\
\int d^{4} x \;  \overline{E}_{p} (x) E_{p'}(x) &=& \left(2\pi\right)^{4} \delta \left( p-p' \right),\\
\int d^{4} p \;  E_{p}(x)  \overline{E}_{p}(x') &=& \left(2\pi\right)^{4} \delta \left( x-x' \right),
\end{eqnarray}
justify a generalized Fourier expansion 
\begin{eqnarray}
	\psi(x) & =&  \int\frac{d^{4}p}{\left(2\pi\right)^{4}}E_{p}(x)\psi(p), 
	\label{fp1}
	\\
	M\left(x',x\right) & = &  \int\frac{d^{4}p}{\left(2\pi\right)^{4}}E_{p}(x')M\left(p,F\right)\overline{E}_{p}(x),\label{eq:mop1}
\end{eqnarray}
where
\begin{equation}\label{F-A}
	F_{\mu \nu} \left( \varphi \right) = k_\mu A_\nu'\left(\varphi \right) - k_\nu A_\mu'\left(\varphi \right)
\end{equation}
is the external EM field tensor,
thereby introducing a $E_p$-representation~\cite{ritus1972radiative,ritus1985quantum}.
By substituting Eqs.~\eqref{fp1} and \eqref{eq:mop1} into the Dyson-Schwinger equation Eq.~\eqref{eq_x}, the latter reduces to a system of algebraic equations:
\begin{equation}
	\left[-\slashed{p}+m+M(p,F)\right]\psi(p)  =  \mathcal{D}(p,F)\psi(p)=0.
	\label{Df}
\end{equation}

Like an arbitrary composition of gamma matrices, $\mathcal{D}(p,F)$ can be decomposed as follows~\cite{berestetskii1982quantum}
\begin{equation}
	\mathcal{D}(p,F)  =  \mathcal{S}+\slashed{\mathcal{V}}+\sigma^{\mu\nu}\mathcal{T}_{\mu\nu}+\slashed{\mathcal{A}}\gamma^{5}+\gamma^{5}\mathcal{P},
	\label{D}
\end{equation}
where $\sigma^{\mu\nu} =  i \left(\gamma^{\mu}\gamma^{\nu} -\gamma^{\nu} \gamma^{\mu} \right)/ 2$. For the crossed field configuration, we have $\mathcal{P}=0$ due to the charge parity conservation. Other coefficients in Eq.~\eqref{D} can be adjusted to the form~\cite{ritus1972radiative} 
\begin{eqnarray}
\mathcal{S} & = & m\;s(p^2,\chi)\\
\mathcal{V}_{\mu} & = & v_{1}(p^2,\chi)p_{\mu}+v_{2}(p^2,\chi)\frac{e^{2}}{m^{4}}F_{\mu\nu}F^{\nu\lambda}p_{\lambda},\label{eq:defV}\\
\mathcal{T}_{\mu\nu} & = & \tau (p^2,\chi)\frac{eF_{\mu\nu}}{m^{2}},\label{eq:defT}\\
\mathcal{A}_{\mu} & = & a(p^2,\chi)\frac{eF_{\mu\nu}^{*}p^{\nu}}{m^{2}},\label{eq:defA}\\
\chi^{2} & = & -\frac{\left(eF^{\mu\nu}p_{\nu}\right)^{2}}{m^{6}}= \xi^2 \frac{p_-^{2}}{m^2},
\end{eqnarray}
where $\chi$ is the Lorentz-invariant quantum dynamical parameter characterizing the interaction of a charged particle with the EM field, $F^*_{\mu \nu} = (1/2) \epsilon_{\mu \nu \alpha \beta } F^{\alpha \beta}$ is the tensor dual to $F^{\mu \nu}$, and $\epsilon_{\mu \nu \alpha \beta }$ is the Levi-Civita symbol.

The solutions to Eq.~\eqref{Df} exist if $\mathrm{det}\left(\mathcal{D}\right)=0$, which reduces to~\cite{ritus1972radiative}:
\begin{equation}
	\label{det_sol}
	d \pm n_D^\mu b_\mu =0,
\end{equation}
where
\begin{eqnarray}
	d & = & \mathcal{S}^{2}+\mathcal{V}^{2}+\mathcal{A}^{2} \\
	&=& m^{2}s^{2} - p^{2}v_{1}^{2} +m^{2}\chi^{2}\left(a^{2} -2v_{1}v_{2}\right),\\
	b_\mu & = & 2\left(\mathcal{S}\mathcal{A}_\mu - 2\mathcal{T}^{*}_{\mu \nu }\mathcal{V}^\nu \right), \nonumber \\
	n_D^\mu b_\mu & = & -2m^{2}\chi\left(sa-2\tau v_{1}\right).
\end{eqnarray}
and 
\begin{eqnarray}
	n^\mu_D &=&  \frac{\mathcal{A}^{\mu} }{a(p^2,\chi) m \chi} = \left( \frac{p_{y}}{p_-}, \mathbf{n} \right), 
	\label{nD}
	\\
	\mathbf{n} &=&  \left\{  0,1,\frac{p_{y}}{p_-}\right\}
\end{eqnarray}
is the unit polarization 4-pseudovector such that $n_D^2 =-1$, $n_D \cdot \mathcal{V} = 0$. Note that it is no more arbitrary and that in the electron proper frame vector $\mathbf{n}$ is directed along the magnetic field (see Appendix~\ref{appendix} and Ref.~\cite{ritus1972radiative}).

Equation \eqref{det_sol} can be recast as the mass shell condition:
\begin{eqnarray}
	p^{2} & = & m_{\uparrow \downarrow}^{2}\equiv m^{2}\left(1+  \mu_{\uparrow \downarrow}\right),\\
	\frac{m_{\uparrow \downarrow}^{2}}{m^2} & = &
	\frac{s^2 + \chi^2 \left( a^2 - 2 v_1 v_2 \right) \pm  2 \chi \left( s a -2 \tau v_1 \right)}{v_1^2}, \;
	\label{ma}
\end{eqnarray}
where the indices $\uparrow,\downarrow$ correspond to the eigenvalues $\sigma=\pm1$ of the projection operator $\gamma^{5}\slashed{n}_D$, i.e. to the two possible spin orientations along and against $n_D$. Thus the radiative corrections result in an electron mass shift which is a complex-valued function of the parameter $\chi$. Since $\mathrm{Im}\left[\mu_{\uparrow \downarrow}\right]<0$, they make the generalized Volkov states decaying~\cite{ritus1972radiative,meuren2011quantum,podszus2021first}.  The origin of their decay is radiation of the electron in the field. In the limit $\alpha \rightarrow 0$ the bare mass shell condition $p^2=m^2$ is recovered.

The positive energy solution to Eq.~\eqref{Df} can be written as~\cite{ritus1972radiative}
\begin{eqnarray}
	\label{fp} \psi (p)  & = &	\psi (\mathbf{p}) \delta(p^0-\mathcal{E}),\\ 
	\psi (\mathbf{p}) & = &
	\overline{\mathcal{D}}\mathcal{L}_{\sigma}w  ,
	\label{U}
\end{eqnarray}
where the delta function sets onto the dressed mass shell $\mathcal{E}  = \sqrt{\mathbf{p}^{2}+m_{\uparrow \downarrow}^{2}}$, $w$ is an arbitrary four-spinor, and
\begin{equation}\label{eq:barD}
\overline{\mathcal{D}}(p,F)  =  \mathcal{S}-\slashed{\mathcal{V}}-\sigma^{\mu\nu}\mathcal{T}_{\mu\nu}+\slashed{\mathcal{A}}\gamma^{5}.
\end{equation}
The  projection operator $\mathcal{L}_{\sigma}$ is given by Eq.~\eqref{Ln}, but with $n=n_D$ (see Appendix~\ref{appendix} and Ref.~\cite{ritus1972radiative}).

Let us now compose an electron wave packet out of the positive energy generalized Volkov solutions~\eqref{fp}. To this end, for simplicity we choose the bispinor $w\left( \mathbf{p} \right)=C\left( \mathbf{p}\right)\times\left\lbrace 0,\,1,\,0,\,0\right\rbrace$, where $C\left( \mathbf{p} \right)$ describes the momentum distribution of the packet. The corresponding wavefunction in the $E_{p}$ representation reads
\begin{eqnarray}
\psi (\mathbf{p}) & = &
\overline{\mathcal{D}}\mathcal{L}_{\sigma}w(\mathbf{p})  =  C(\mathbf{p})U_{\sigma}(\mathbf{p}) ,
\label{U}
\end{eqnarray}
where
\begin{eqnarray}
	U_{\sigma}(\mathbf{p}) & = & m \left(\begin{array}{c}
	Q_\sigma + i \sigma K_\sigma  \\
	P_\sigma - K_\sigma + \sigma v_1 \frac{p_y}{m} \\
	Q_\sigma - v_1 \frac{p_{xy} - i \sigma p_z}{m}    \\
	- P_\sigma + v_1 \frac{p_z + i \sigma p_x}{m}
	\end{array}\right),
\\
P_\sigma & = &  -a \xi  \frac{p_y}{m} - v_2  \xi \chi + 2 \sigma \tau \xi  -  \sigma \frac{s p_y}{\mathcal{E} - p_z} , \nonumber \\
Q_\sigma & = & - 2 i  \tau \xi  +  i \sigma v_2 \xi \chi  + \sigma v_1  \frac{p_{y}}{\mathcal{E} - p_z} \frac{p_{xy}}{m} , \nonumber \\
K_\sigma & = & -s +  v_1 \frac{\mathcal{E}}{m} - \sigma a \chi , \nonumber \\
p_{xy} & = & p_{x}-ip_{y}. \nonumber
\end{eqnarray}

Up to this point the expression for $\psi(\mathbf{p})$ is formally exact. However, the scalar functions $s$, $v_1$, $v_2$, $\tau$ and $a$ are known only in a one-loop approximation~\cite{ritus1972radiative}. On the mass shell (with the same accuracy $p^2 = m^2$) they depend solely on $\chi$ and read:
\begin{eqnarray}
s & = & 1+\frac{\alpha}{\pi}\int_{0}^{\infty}\frac{du}{\left(1+u\right)^{2}}f_{1}\left(\lambda \right), \label{s} \\
v_{1} & = & -1-\frac{\alpha}{2\pi}\int_{0}^{\infty}\frac{du}{\left(1+u\right)^{3}}f_{1}\left(\lambda \right),\\
v_{2} & = & -\frac{\alpha}{2\pi\chi^{2}}\int_{0}^{\infty}\frac{du}{\left(1+u\right)^{3}} f_2 (u,\lambda), \\
\tau  & = & -\frac{\alpha}{2\pi\chi}\int_{0}^{\infty}\frac{du}{\left(1+u\right)^{2}} \lambda^{-1/2}f\left(\lambda \right),\\
a & = & \frac{\alpha}{2\pi\chi}\int_{0}^{\infty}\frac{(2+u)du}{\left(1+u\right)^{3}} \lambda^{-1/2}f\left(\lambda \right),\label{f2}\\
f_2 (u,\lambda) &=& u f_{1}\left(\lambda \right) -\left(u^{2}+2u+2\right) \frac{1}{\lambda} \frac{df\left(\lambda \right)}{d\lambda},  \nonumber \\
\lambda &=& \frac{u^{2/3}}{\chi^{2/3}}, \nonumber
\end{eqnarray}
in terms of the functions $f(\lambda)$ and $f_{1}(\lambda)$ defined by~\cite{ritus1972radiative}
\begin{eqnarray}
	f(\lambda) & = & i\intop_{0}^{\infty}d\zeta\exp\left(-i\lambda  \zeta-i\frac{\zeta^{3}}{3}\right),\\
	f_{1}(\lambda) & = & \intop_{\lambda}^{\infty}dx\left[f(x)-\frac{1}{x}\right] \nonumber  \\ 
	&=& \intop_{0}^{\infty}\frac{d\zeta}{\zeta}\exp\left(-i\lambda \zeta\right) \left[\exp\left(-i\frac{\zeta^{3}}{3}\right)-1\right].
	\end{eqnarray}
In the limit $\alpha \rightarrow 0$ we have $s= - v_1 = 1$, $a=\tau=v_2 =\mu_{\uparrow \downarrow} = 0$ and $p^2 = m^2$, so that $\mathcal{D} = \mathcal{D}_0^{(+)}$, $\overline{\mathcal{D}} = \mathcal{D}_0^{(-)}$ and the Dyson-Schwinger equation reduces to the Dirac equation. 

In a strong field limit $\chi\gg1$, the integrals in Eqs.~\eqref{s}-\eqref{f2} are formed at small argument of $f(\lambda)$,
$f'(\lambda)$, and $f_{1}(\lambda)$. For $\lambda \ll 1$, these functions are approximated as $f(\lambda) = c_1+O(\lambda)$, $f'(\lambda) = c_2+O(\lambda)$ and $f_{1}(z) = \ln\left(\lambda\right)+c_{3}+O(\lambda)$, respectively, where
\begin{eqnarray}
c_{1} & = & \frac{\Gamma (1/3)}{3^{2/3} 2}\left(1+i\sqrt{3}\right), \nonumber \\
c_{2} &=& \frac{3^{2/3}\Gamma\left(2/3\right)}{6}\left(1-i\sqrt{3}\right), \nonumber \\
c_{3} & = & \frac{2}{3}\gamma+\frac{1}{3}\ln3+i\frac{\pi}{3}. \nonumber
\end{eqnarray}
After integrating over $u$, for $\chi\gg 1$ the scalar parameters acquire the form 
\begin{eqnarray}
	 \overline{s} & \equiv & s-1 = \frac{\alpha}{\pi}\left(-\frac{2}{3}\ln\chi+c_{3}\right)+O\left(\chi^{-2/3}\right),
	\label{s-1} \\
	\tau & = & -\frac{\alpha c_{1}}{3\sqrt{3}\chi^{2/3}}+O\left(\chi^{-4/3}\right),\\
	a & = & \frac{5\alpha c_{1}}{9\sqrt{3}\chi^{2/3}}+O\left(\chi^{-4/3}\right),\label{eq:a}\\
	\overline{v}_1 & \equiv & v_{1}+1 =  \frac{\alpha}{2\pi}\left(\frac{1}{3}\ln\chi+c_3'\right)+O\left(\chi^{-2/3}\right),\\
	v_{2} & = & \frac{\alpha c_{2}}{\chi^{4/3}}\frac{14}{9\sqrt{3}}+O\left(\chi^{-2}\ln \chi \right),
	\label{v2}
\end{eqnarray}
where $c_3'=1/3-c_{3}/2$. Accordingly, the mass shift given by Eq.~\eqref{ma} takes the form \cite{BaierKatkov,ritus1985quantum}:
\begin{equation}
\begin{split}
\frac{\mu_{\uparrow \downarrow}}{2}  & \simeq   s_1 + v_1  + v_2 \chi^2 \pm \chi (a - 2 \tau) + O(\alpha^2)
\\
& \simeq   \frac{7 \Gamma (2/3) \left( 1 - i \sqrt{3} \right) }{27 \sqrt{3}} \alpha  \left( 3 \chi \right)^{2/3} 
\\
&\mp  \frac{\Gamma (1/3) \left( 1 + i \sqrt{3}\right) }{ 54 \sqrt{3}  } \alpha \left( 3 \chi \right)^{1/3} \\
&\quad\quad\quad\quad\quad\quad\quad\quad + O\left(\alpha \ln \chi, \alpha^{2} \right).
\end{split}
\end{equation}
Notably, the correction $\mathrm{Re}\,\mu_{\uparrow\downarrow}$ to the electron mass is negative for the spin up state $\sigma=+1$ and positive for the spin down state $\sigma=-1$. It is associated with the anomalous magnetic moment of the electron. In accordance with Eq.~\eqref{nD}, the electron polarization vector has a component along the magnetic field, thereby providing a reduction of the electron energy $\propto - \mu' B $, where $\mu'$ is the anomalous magnetic moment.

\section{Wave packet dynamics in coordinate space}
\label{sec3}
Transition to the coordinate space is accomplished by means of~Eq.~\eqref{fp1}. In virtue of~Eqs.~\eqref{fp1}, \eqref{fp} and \eqref{U}, the wavefunction reads
\begin{eqnarray}
	\psi (x) & = & \int\frac{d\mathbf{p}}{\left(2\pi\right)^{3}}\left.\vphantom{\frac12}E_{p}(x)\psi (\mathbf{p}) \right|_{p^0=\mathcal{E}}
	\nonumber
	\\
	& = & \int\frac{d\mathbf{p}}{\left(2\pi\right)^{3}} E_{p}(x)C(\mathbf{p}) U_{\sigma}(\mathbf{p}).
   	\label{psi1}
\end{eqnarray}

We assume that the wave packet has a Gaussian shape and is narrow in the momentum space:
\begin{equation}
	C(\mathbf{p}) =  N \exp\left[-\frac{\left(p_{z}+\varepsilon\right)^{2}}{2\Delta_{z}^{2}}-\frac{p_{\bot}^{2}}{2\Delta_{\bot}^{2}}\right],
	\label{cp}
\end{equation}
where $\varepsilon \gg  \left\{ \Delta_{z},\Delta_{\perp}\right\} \gg\left|m_{\uparrow \downarrow}\right|$, $p^2_{\bot} = p_x^2 + p_y^2 $ and $N$ is the normalizing coefficient. This corresponds to an ultrarelativistic electron initially moving to the negative direction of the axis~$z$ with the energy $\simeq \varepsilon $ much greater than the momentum spread of the  packet. The integrand in Eq.~\eqref{psi1} can be represented in an exponential form:
\begin{eqnarray}
 E_{p}(x)C(\mathbf{p}) U_{\sigma}(\mathbf{p}) & = &  G (\mathbf{p}) e^{- \Phi \left( \mathbf{p} \right)},
\label{Eppsip}
 \\
 \Phi \left( \mathbf{p} \right) & = &  \frac{\left(p_{z}+\varepsilon\right)^{2}}{2\Delta_{z}^{2}} +\frac{p_{\bot}^{2}}{2\Delta_{\bot}^{2}} - i \Phi_V  . 
\label{phi1}
\end{eqnarray}

It is convenient to introduce a deviation $\overline{p}_z =p_{z}+\varepsilon$ of the $z$-component of the momentum from the central value. It follows from Eq.~\eqref{cp} that the main contribution to the integral in Eq.~\eqref{psi1} comes from the momentum volume $ \left|\overline{p}_z \right|  \lesssim \Delta_{z}$, $p_\perp \lesssim \Delta_{\perp}$. Thus, introducing $\Pi_{\uparrow \downarrow}^{2}=p_{x}^{2}+p_{y}^{2}+m_{\uparrow \downarrow}^{2}$, we can use the  expansions:
\begin{eqnarray}
	p^0 &=& \mathcal{E} = \sqrt{p_{z}^{2}+\Pi_{\uparrow \downarrow}^{2}} 
	\nonumber
	\\
	& = & \varepsilon - \overline{p}_z +\frac{\Pi_{\uparrow \downarrow}^{2}}{2\varepsilon } + \frac{\overline{p}_z \Pi_{\uparrow \downarrow}^{2}}{2\varepsilon^2} + \ldots ,
	\label{ser1}
	 \;\; \\
		\chi & = & \chi_{0}\left(1-\frac{\overline{p}_z}{\varepsilon} 
	 +  \frac{\Pi_{\uparrow \downarrow}^{2}}{4\varepsilon^2}+  \frac{\overline{p}_z \Pi_{\uparrow \downarrow}^{2}}{4\varepsilon^{3}} + \ldots \right),\\
	\frac{2\varepsilon}{p_-} & = & 1+\frac{\overline{p}_z}{\varepsilon }
		+ \frac{4\overline{p}_z^2-\Pi_{\uparrow \downarrow}^{2}}{4\varepsilon^2}
	+ \frac{4\overline{p}_z^3-3\Pi_{\uparrow \downarrow}^{2}\overline{p}_z }{4\varepsilon^3}+\ldots ,\;\;
	\label{ser3}
	\end{eqnarray}
where $\chi_{0}=2 \xi \varepsilon / m $. 
Substituting Eqs.~\eqref{ser1}-\eqref{ser3} into Eq.~\eqref{phi1}, we obtain:
\begin{eqnarray}
\Phi (\mathbf{p}) & = &  \frac{p_\bot^2}{2 D_\bot^{2}} + \frac{\overline{p}_z^2}{2 D_{z}^{2} } - i  \left (\overline{p}_z - \varepsilon \right) (t+z) - i \left(\mathbf{p}_{\perp} \mathbf{r}_{\perp}\right) 
\nonumber
\\
& + &  i  t \frac{m_{\uparrow \downarrow}^{2}}{2 \varepsilon} - i \frac{\Lambda \varphi^2 p_x}{4 m } + i \frac{\Lambda^ 2 \varphi^3 }{12 m } \left( \varepsilon + \overline{p}_z \right )  \nonumber
\\
& - &  i \frac{m_{\uparrow \downarrow}^{2}\Lambda^ 2 \varphi^3 }{48 m \varepsilon}   + R, \;\;
\label{phip}
\end{eqnarray}
where
\begin{eqnarray}
\frac{1}{D_z^2} & = & \frac{1}{\Delta_z^2} + i \frac{\Lambda^ 2 \varphi^3 }{6 m \varepsilon} , \\
\frac{1}{D_\bot^2} &=& \frac{1}{\Delta_\bot^2} +  i  \frac{t}{ \varepsilon}
- i \frac{\Lambda^ 2 \varphi^3 }{24 m\varepsilon} ,
\end{eqnarray}
and $R$ stands for the remainder terms stemming from the last shown and the omitted higher-order terms in the series \eqref{ser1}-\eqref{ser3}.
Since $\Phi$ stands in the argument of the exponential, the remainder terms can be neglected if $R \ll 1$. In virtue of Eqs.~\eqref{phase}~and~\eqref{ser1}-\eqref{ser3} this condition is expanded as follows:
\begin{equation}
	\Delta_z \left\{ \frac{ \Delta_\bot \Lambda \varphi^2 }{4 m \varepsilon } ,\; \frac{  \left(4\Delta_z^2 + 3\Delta_\bot^2 \right)  \Lambda^2 \varphi^3}{24 m \varepsilon^2 }  , \; \frac{\Delta_\bot^2 t}{2 \varepsilon^{2} }\right\} \ll 1 . 
	\label{neq1}
\end{equation}
Eq.~\eqref{neq1} points to an upper bound on duration $t$ for which the developed approximation remains valid. 

By the same token, we expand the pre-exponential factor in Eq.~\eqref{Eppsip} to first order in powers of $\alpha$, $\overline{p}_z/\varepsilon$ and $p_\perp / \varepsilon$:
\begin{eqnarray}
G(\mathbf{p}) & = &  N \varepsilon \left[ G_0(\mathbf{p}) + \Lambda \varphi G_\Lambda (\mathbf{p}) \right]   
\nonumber
\\
&+&  O \left[ \left(\frac{m}{\varepsilon}\right)^2, \left(\frac{\overline{p}_z}{\varepsilon}\right)^{2},\left(\frac{p_{\perp}}{\varepsilon}\right)^{2},\alpha^{2} \right], \;\;\;\;\;\;
\label{g}
\end{eqnarray}
where
\begin{eqnarray}
G_0 (\mathbf{p}) & = &  \left(\begin{array}{c}
2 Q_1   -i \sigma  \left( P_1 - \frac{\overline{p}_z}{\varepsilon}  \right) \\
P_1 - \frac{p_z}{\varepsilon} +\sigma \left( 2 i Q_1- \frac{p_y}{\varepsilon}  \right)  \\
\frac{p_{xy}}{\varepsilon}  + i\sigma \left(P_2-  \frac{\overline{p}_z}{\varepsilon}\right)  \\
P_2 - \frac{\overline{p}_z}{\varepsilon}  - i\sigma \frac{p_x}{\varepsilon}  
\end{array}\right),
\\
G_\Lambda (\mathbf{p})  & = & -\frac{ 1}{2} \left(\begin{array}{c}
Q_2 + i \sigma \left(Q_1  - \frac{p_{xy}}{2 \varepsilon} \right) \\
 - Q_1  + i \sigma Q_2+\frac{p_{xy}}{2\varepsilon} \\
Q_2 + i\sigma \left(Q_1- \frac{p_{xy}}{2\varepsilon} \right) \\
Q_1  - i\sigma Q_2  - \frac{p_{xy}}{2\varepsilon}    
\end{array}\right),
\\
P_1 &=& \frac{m}{\varepsilon} -  v_1 - 2 v_2 \xi^2  , \\
P_2 &=&  -v_1 + 2 v_2 \xi^2  ,\\
Q_1 &=& -i a \xi, \\
Q_2 &=& \frac{m}{2 \varepsilon} - v_1   .
\end{eqnarray}

After integrating in Eq.~\eqref{psi1} over the momentum, we obtain the wavefunction in the coordinate space: 
\begin{eqnarray}
\psi (x) & =&  N \varepsilon D_{z} D_\bot^2  \left( 2\pi \right)^{3/2} \left[ G_0(x) + \Lambda \varphi G_\Lambda (x) \right]  e^{- \Phi (x)}
\nonumber
\\
&+& O \left[ \left(\frac{m}{\varepsilon}\right)^2, \left(\frac{\Delta_z}{\varepsilon}\right)^{2},\left(\frac{\Delta_\bot}{\varepsilon}\right)^{2},\alpha^{2} \right], 
	\label{phix}
\end{eqnarray}
where 
\begin{eqnarray}
		\Phi (x) & = & \frac{D_{z}^{2} Z^{2} + D^2_\bot \left( y^{2}+X^{2} \right)}{2}
		\nonumber 
		\\ 
		&+& i \frac{m_{\uparrow \downarrow}^{2}}{2\varepsilon} T 
		+ i \frac{\Lambda^ 2 \varphi^3 \varepsilon}{12 m }
		 + i\varepsilon (t+z), \;\;\;\;
		\label{phi0} \\
	G_0 (x) & = & \left(\begin{array}{c}
	2 Q_1 - \sigma \left( i P_1 + \frac{D_z^2 Z}{\varepsilon}  \right) \\
	P_1 - i \frac{D_z^2 Z}{\varepsilon} + i\sigma \left(  2  Q_1 - \frac{D_\bot^2 y}{\varepsilon}  \right)  \\
	i \frac{D_\bot^2 \left( X - iy \right)}{\varepsilon}  +\sigma  \left( i P_2 + \frac{D_z^2 Z}{\varepsilon} \right)  \\
	P_2 - i \frac{D_z^2 Z}{\varepsilon}  + \sigma \frac{D_\bot^2 X}{\varepsilon}  
	\end{array}\right),
	\\
	G_\Lambda (x)  & = & -\frac{1}{2} \left(\begin{array}{c}
	Q_2 +\sigma  \left[i Q_1+ \frac{D_\bot^2 \left( X - iy \right)}{2\varepsilon}  \right] \\
	 - Q_1  + i\sigma Q_2 +i \frac{D_\bot^2 \left( X - iy \right)}{2\varepsilon}
	\\
Q_2 +\sigma  \left[ i Q_1 + \frac{D_\bot^2 \left( X - iy \right)}{2\varepsilon} \right] 
 \\
	Q_1 - i \sigma Q_2  - i \frac{D_\bot^2 \left( X - iy \right)}{2\varepsilon}    
	\end{array}\right),
\end{eqnarray}
and
\begin{eqnarray}
	Z & = & t+z- \frac{\Lambda^ 2 \varphi^3 }{12 m },
	\label{Z}
	\\
	X & = & x + \frac{\Lambda  \varphi^2 }{4 m },
	\\
	T & = & t - \frac{\Lambda^2  \varphi^3 }{24m}
	\label{T}. 
\end{eqnarray}
One can see from Eq.~\eqref{phi0} that the wavefunction is localized in the time-space domain $ Z^2  \lesssim 1/\Delta_z^2$, $X^2 + y^2 \lesssim 1/\Delta_{\perp}^2$. 

Next we turn to the probability density
\begin{equation}
	\rho(x)  =  \psi^{\dagger}(x)\psi(x).
	\label{density}
\end{equation}
In virtue of Eqs.~\eqref{phix}-\eqref{density} we obtain
\begin{eqnarray}
	\rho (x) & = & 
	C_\rho |D_z|^2 |D_\perp|^4 \left( \rho_{0} + \alpha \rho_{\alpha}  \right) e^{ -\Phi_\rho - W_r T}, 
	\label{rho_WT}\\
	\Phi_\rho (x) & = &  \mathrm{Re}\left[ D_z^2 \right] Z^2 + \mathrm{Re}\left[ D_\bot^2 \right] \left(y^{2}+X^{2}\right),
	\label{phirho}
	\\
	\rho_0 & = & \frac{1}{\varepsilon}\left\lbrace 2 \mathrm{Im}\,D_z^2 Z +\frac{1}{2} \mathrm{Im}\,D_\perp^2 X \Lambda\varphi \right.\\
	\nonumber
	 &&+\left(1+\frac{\Lambda^2\varphi^2}{4}\right)\left[\varepsilon+m \vphantom{D_\perp^2 }\right.\\
	\nonumber
	&& \quad\quad\quad\quad\quad\quad\left.\left.+\sigma\left(\mathrm{Re}\,D_\perp^2 X + \mathrm{Im}\,D_\perp^2 y\right)\right] \vphantom{\frac{1}{2}}\right\rbrace
\\
\rho_{\alpha} (x) & = & \frac{4  + \Lambda^2\varphi^2}{2}   \mathrm{Re}\,g_{1} + \Lambda\varphi \mathrm{Im}\,g_2 , 
\\
g_{1} & = & - \frac{\overline{v}_1}{\alpha} +\sigma \frac{\chi m}{2 \varepsilon}\frac{a}{\alpha}, \\
g_{2} & = & - \frac{\chi m}{2 \varepsilon}\frac{a}{\alpha} + \sigma \frac{\chi^2 m^2}{2 \varepsilon^2}\frac{v_2}{\alpha}
\end{eqnarray}
where $C_\rho = N^2 (2\pi)^3 4\varepsilon^2$ is an overall constant, and 
\begin{equation} 
W_r = - \mathrm{Im}\left[ \frac{m_{\uparrow \downarrow}^{2}}{ \varepsilon} \right] \simeq
\frac{14 \Gamma \left( 2/3\right)}{3^{7/3}} \alpha \chi^{2/3} \frac{m^2}{\varepsilon}
\label{wr}
\end{equation}
is the total probability of photon emission \cite{BaierKatkov,ritus1985quantum}.
The term $-W_r T$ in the exponential describes damping of the probability density due to the leakage of a part of the wavefunction to a many-particle sector combining electron with the emitted photons \cite{ritus1972radiative,meuren2011quantum,podszus2021first}. Here we focus only on the remaining part of the wavefunction with no emitted photons, which gets damped.   

Note that for $\chi \gg 1$ (implying also the field subcritical $\xi<1$ but $\varepsilon\gg m$), using Eqs.~\eqref{s-1}-\eqref{v2} we can estimate the coefficients  $g_{1}$, $g_{2}$ as follows:
\begin{eqnarray}
\mathrm{Re}\, g_{1} & \simeq &  \frac{\ln \chi}{6 \pi} , \\
\mathrm{Re}\, g_{2} & = &  O\left( \frac{m^{2/3} \xi^{1/3}}{\varepsilon^{2/3}}\right)\ll \mathrm{Re}\, g_{1}.
\end{eqnarray}

Assuming the probability density is normalized at $t=0$ by
\begin{equation}
\int\limits \rho (t=0,\mathbf{r})\,d\mathbf{r} = 1,
\label{norma}
\end{equation}
one can recover the overall normalization constant $N$. Assuming for simplicity that $\Lambda\varphi \ll 1$, we have $z\approx -t$ and $\varphi \approx 2mt$. In this approximation we obtain
\begin{equation}
	N^2\approx \frac{1}{32 \pi^{9/2} \Delta_z \Delta_\perp^2 (m+\varepsilon)^2}\left(1+\frac{m}{\varepsilon}-2\alpha \mathrm{Re}\,g_1\right).
\end{equation}
Finally, the probability density takes form
\begin{eqnarray}
\rho (x) & = & 
 \left(\rho_{0}+\alpha\rho_{\alpha}\right) \frac{\varepsilon^2 \Delta_z \Delta_\bot^2 e^{-\Phi_\rho - W_r T}} {\pi^{3/2} \left( \varepsilon^2 + \Delta_{\bot}^4 t^2 \right) }
 \nonumber 
\\
&+& O \left( \frac{m^2}{\varepsilon^2}, \frac{\Delta^{2}}{\varepsilon^{2}}, \alpha^{2}\right), 
\label{rho1}
\\
\Phi_\rho (x) & = &  \Delta_z^2 (z+t)^2 + \frac{\Delta_{\bot}^{2}\varepsilon^{2} \left(y^{2}+X^{2}\right)}{\varepsilon^{2}+t^{2}\Delta_{\bot}^{4}},
\\
\rho_0 & = & 1  +\sigma \Delta_\perp^2 \frac{\varepsilon X - \Delta_\perp^2 t y}{\varepsilon^2 + \Delta_{\bot}^4 t^2},\\
\rho_\alpha & =&  -\left(\frac{4m}{\varepsilon} +2 \sigma \Delta_\perp^2 \frac{\varepsilon X - \Delta_\perp^2 t y}{\varepsilon^2 + \Delta_{\bot}^4 t^2} \right) \mathrm{Re}\, g_{1}.
\label{rho4}
\end{eqnarray}
It follows that the center of the wave packet propagates along the trajectory $X=y=Z=0$ as might be expected. Its longitudinal width (along $z$-axis) is $\sim 1/\Delta_z $. As for its transverse width (along $x$- and $y$-axes), it expands as $\sqrt{1/\Delta_\bot^2+\Delta_\bot^2 t^2/\varepsilon^2}$.

\section{Expectation value of the Dirac spin operator}
\label{sec4}
An expectation value of the Dirac spin operator for the wave packet is given by
\begin{equation}
\left\langle \mathbf{S} \right\rangle = \frac{1}{2}\int\limits d \mathbf{r}  \; \psi^{\dagger}(x) \boldsymbol{\Sigma}   \psi(x), 
\label{spin}
\end{equation} 
where $\boldsymbol{\Sigma} = \gamma^0 \boldsymbol{\gamma} \gamma^5 $ is the three dimensional spin operator \cite{berestetskii1982quantum}.
Let us pass to the $E_p$-representation, 
\begin{eqnarray}
\left\langle \mathbf{S} \right\rangle =\frac{1}{2}\int\limits d \mathbf{r} \frac{d \mathbf{p}}{(2\pi)^3} \frac{d \mathbf{p}'}{(2\pi)^3} \, \psi^{\dagger}(\mathbf{p}' ) E_{p'}^{\dagger}(x) \boldsymbol{\Sigma} E_p (x ) \psi(\mathbf{p}),
\label{spin1}
\end{eqnarray}
where $E_{p,p'}(x)$ are taken on the mass shell $p^0=\mathcal{E}$ [see Eqs.~\eqref{psi1} and \eqref{Eppsip}].
After consequent integration over $\mathbf{r}_\bot$ and $\mathbf{p}'_\bot$, Eq.~\eqref{spin1} acquires the form 
\begin{eqnarray}
\left\langle \mathbf{S} \right\rangle &=& \frac{1}{2(2\pi)^4} \int_{-\infty}^{+\infty} dz\, d \mathbf{p}\, d \mathbf{p'} \, G^{\dagger}(\mathbf{p'}) \boldsymbol{\Sigma}   G(\mathbf{p}) 
\nonumber
\\
& & \quad\quad\quad\quad \times \delta \left( \mathbf{p_\bot} - \mathbf{p'_\bot}  \right) \exp \left[ - \Phi_S (\mathbf{p},\mathbf{p}')   \right]
\\
& = & \frac{N^2\varepsilon^2}{2(2\pi)^4} \int_{-\infty}^{+\infty} d z\, d p_z\, d p^{\prime}_z\, d\mathbf{p_\bot} \; \left(\mathbf{B}_0 + z \mathbf{B}_1 + z^2 \mathbf{B}_2  \right) 
\nonumber \\
& & \quad\quad  \times\exp \left[ i z (p_z - p'_z) \right],
\end{eqnarray}
where
\begin{eqnarray}
\mathbf{B}_0 & = & e^{-\Phi_S} B_G^\dagger (\mathbf{p}') \boldsymbol{\Sigma}  B_G(\mathbf{p}) , \\
\mathbf{B}_1 & = &  - e^{-\Phi_S} \Lambda \left( G^\dagger_\Lambda (\mathbf{p}') \boldsymbol{\Sigma}  B_G (\mathbf{p} ) \right.
\nonumber
\\
&&\left. \quad\quad\quad\quad\quad + B_G^ \dagger (\mathbf{p}' ) \boldsymbol{\Sigma} G_\Lambda (\mathbf{p} ) \right), \\
\mathbf{B}_2 & = & e^{-\Phi_S} G^\dagger_\Lambda (\mathbf{p}' ) \boldsymbol{\Sigma}   G_\Lambda (\mathbf{p}) , \\
B_G(\mathbf{p}) & = & G_0 (\mathbf{p}) + \Lambda t G_\Lambda (\mathbf{p}) , \\
\Phi_S (\mathbf{p},\mathbf{p}')  & = & 
 \Phi(\mathbf{p}) + \Phi^*(\mathbf{p}'),
\end{eqnarray}
and it is implied that $\mathbf{p}_\perp'=\mathbf{p}_\perp$. By integrating over $z$, we next obtain
\begin{eqnarray}
\left\langle \mathbf{S} \right\rangle &=& \frac{N^2\varepsilon^2}{2(2\pi)^3} \int_{-\infty}^{+\infty} d p_z d p'_z d\mathbf{p_\bot}    
\nonumber
\\
&& \times \left[ \vphantom{\frac{d^2}{d\overline{p}_z^{\prime 2}}} \mathbf{B}_0  \right. +\left. i \mathbf{B}_1 \frac{d}{dp'_z}   -   \mathbf{B}_2  \frac{d^2}{dp_z^{\prime 2}}   \right]\delta (p_z - p'_z)
\nonumber
\\
&=& \frac{N^2\varepsilon^2}{2(2\pi)^3} \int_{-\infty}^{+\infty} d\mathbf{p}   \left( \mathbf{B}_0   
-  i  \frac{\partial  \mathbf{B}_1}{\partial p'_z}  
\right. 
\nonumber
\\
&& \quad\quad\quad\quad\quad\quad\quad\quad - \left. \left.  \frac{\partial ^2 \mathbf{B}_2}{\partial p^{'2}_z}  \right)  \right|_{p'_z  = p_z}.
\end{eqnarray}

Further calculation proceeds the same way as was used to derive Eqs.~\eqref{rho1}--\eqref{rho4}. We thus obtain that to the first order in $\alpha$, $\Delta_{z,\perp}/\varepsilon$ and $m/\varepsilon$, the only non-vanishing component is 
\begin{eqnarray}
	\left\langle  S_y \right\rangle & \approx & \sigma  \frac{m}{2\varepsilon} e^{- W_r t} \left( 1 + \alpha  \mathrm{Re}\,g_{3} \right) ,
\end{eqnarray}
where 
\begin{eqnarray}
g_{3} & = & \sigma \chi \frac{a}{\alpha}  - \frac{m}{\varepsilon} \chi^2 \frac{v_2}{\alpha} - g_{1 }.
\end{eqnarray}
Note that for $\chi \gg 1$ (also implying  the field subcritical $\xi<1$ but $\varepsilon\gg m$), using Eqs.~\eqref{s-1}-\eqref{v2} we have
\begin{eqnarray}
\mathrm{Re}\, g_{3} & \simeq & \sigma  \frac{2 \chi^{1/3}}{\Gamma \left( 2/3 \right) 3^{11/3}} -  \frac{\ln \chi}{6 \pi}.
\end{eqnarray}
Let us stress that for the considered packet all the components of spin are not vanishing precisely but in the adopted approximations the dominant one is those parallel to the magnetic field while the others show up in higher orders. 

\section{Discussion}
\label{sec5}

According to classical electrodynamics an electron trajectory in the crossed fields can be written in a parametric form  (see, for example, Ref.~\cite{landau2013classical}) 
\begin{eqnarray}
x & = & - \frac{p^2_x }{2 \xi m^2 p_{-} },
\label{tr1}
\\
y & = & - \frac{p_x p_y }{\xi m^2 p_{-} },
\\
z & = &  \frac{p_x }{2 \xi m^2 } \left( 1 - \frac{m^2 + p_y^2 }{p_{-}^2}  \right) - \frac{p^3_x }{6 \xi m^2 p_{-}^2 },
\\
t & = & -\left( \frac{m^2 + p_y^2 }{p_{-}^2}  + 1 \right)  \frac{p_x }{2 \xi m^2 }  - \frac{p^3_x }{6 \xi m^2 p_{-}^2 },
\label{tr4}
\end{eqnarray} 
where $p_- = \varepsilon - p_z$, $\varepsilon$ is the electron energy and initial conditions $\mathbf{r}(t=0) = 0$, $p_x (t=0) = 0$ are assumed. 
In the ultrarelativistic limit $\varepsilon \gg p_\bot \gg m$ ($p_- \simeq 2 \varepsilon$) and for short-term dynamics of the electron dynamics $t \ll \sqrt{3} \chi / \left( m \xi^2 \right) \equiv t_{cl}$  the electron trajectory can be rewritten in the explicit form as follows
\begin{eqnarray}
x & = & - \Lambda m t^2,
\label{tr51}
\\
y & = & 0,
\\
z & = & - t \left( 1 -  \frac{2 \Lambda^2 t^2 m^2 }{3 } \right),
\\
\tau_e & = & -\frac{p_x }{\xi m p_{-} } = \frac{m}{\varepsilon} t \left(1  -\frac{ \Lambda^2 t^2 m^2 }{3 } \right),
\label{tr54}
\end{eqnarray} 
where $\tau_e$ is the proper time of the electron. It is seen from Eqs.~\eqref{tr51}-\eqref{tr54} that the classical trajectory is in agreement with the trajectory of the electron packet center given by Eqs.~\eqref{Z}-\eqref{T} where $\varphi \simeq 2 m t$ for $\Lambda \varphi \ll 1$. Moreover $T$ is proportional to the classical proper time of the electron.

Let us assume for simplicity that $\Delta_z = \Delta_\bot = \Delta$. We can rewrite the mentioned above validity conditions as 
\begin{equation} 
t \ll  \frac{\chi}{\xi^2 \Delta} \min \left\{ 1, \left( \frac{4 \chi} {7}\right)^{1/3}, \frac{2 \chi m^2 } {\Delta^2}, \frac{\Delta} {2 m}  \right\}.
\label{neq3}
\end{equation}
Assuming that $\chi>1$, $\Delta \gg m$ and $\xi \ll 1$, the second and the fourth terms in RHS of Eq.~\eqref{neq3} are greater than the others and therefore can be omitted. If we are interested in maximizing of $W_r t $, then for the parameters of interest ($\chi > 1$, $\xi <1$, $\Delta > m$) the third term  can also be neglected. Therefore the time interval within which our approximations are valid can be presented as follows
\begin{equation}
t \ll t_m \equiv \frac{\chi}{\xi^2 \Delta} \ll t_{cl}. 
\label{tm}
\end{equation} 

As an example, let us consider the electron wave packet with energy $400$~GeV propagating perpendicular to the magnetic and electric field $E=B=7 \times 10^{-4} E_{cr}$ so that $\chi = 1096$. We set the longitudinal and transversal energy spreads of the packet to $\Delta_z = \Delta_{\bot} = \Delta = 1$~GeV. 
Our approach is accurate for the time $t \ll t_{m}  \simeq 4.6 \times 10^6 \hbar / (mc^2)$.
For $t=t_{m}$ the suppression of the probability density for non-emitting electron states is significant: $\exp (- W_r t_{m}) \simeq 0.2 $.
It follows from Eq.~\eqref{tm} that $t_m$ increases with decreasing the electron packet momentum spread. For example, $\exp (- W_r t_m) \simeq 10^{-144} $ for $\Delta=5$~MeV that is the non-emitting state of the electron is completely damped at $t \simeq t_m$. Therefore, despite the fact that we use short-term approximation, the derived solutions are still valid for rather long time interval within which the non-emitting electron states can be completely damped. 
The contribution of the radiative corrections to the probability density is very small $\alpha \rho_\alpha \simeq \alpha (m^4 / \varepsilon^2 \Delta^2 ) \chi^2 \ln \chi / (6 \xi) \simeq 2.5 \times 10^{-7}$. However the contribution of the radiative corrections to the averaged spin is much higher $|1 - \left\langle  S_y \right\rangle / \left\langle  S_y (\alpha = 0) \right\rangle |   \simeq  0.03 \alpha \chi^{1/3} \simeq 2 \times 10^{-3}$. Thus the spin variables is more beneficial to detect the effect of the radiative corrections than the probability density. 

It is interesting to note that the expected damping factor $\exp(-W_r t)$ in Eq.~\eqref{rho_WT} for the wave packet density evolution is enhanced by the term that can be associated to the one $\propto A^2$ in the phase $\Phi_V$ of the Volkov function [see~Eqs.~\eqref{fv} and \eqref{phase}]:
\begin{equation}
W_r T = W_r \left( t - \frac{\Lambda^2 \varphi^3}{24m} \right) \simeq 
W_r t \left( 1-\frac{ \xi^2 m^4}{3 \varepsilon^2 } t^2  \right).
\label{dfactor}
\end{equation}  
It follows from Eqs.~\eqref{dfactor} and \eqref{tr54} that the damping factor is proportional to the proper time of the electron. The term proportional to $\xi^2$  is much smaller than the main term for $t < t_m$: $\xi m^2 t_m / \varepsilon \ll 1$. If it is neglected then damping rate is equal to $W_r$.

To conclude, the evolution of an electron wave packet in a strong constant crossed electromagnetic field is studied with taking into account the radiative corrections caused by the interaction of the electron with the vacuum fluctuations. The dynamics of the wave packet obeys the Dyson-Schwinger equation, which can be formally solved exactly in the $E_p$-representation. We derive the approximate solution in the configuration space at one-loop level. For this, we define the initial packet in the $p$-space and assign it a Gaussian shape. We assume that the packet width in the momentum space is small compared to the particle energy. The obtained result corresponds to evolution of a wave packet without real photon emission. The time of the solution validity is restricted from above. 
The radiative corrections modify the structure of the electron wave function, in particular, result in wave packet damping. The expectation value of the Dirac spin operator is also calculated. The radiative corrections make greater contribution to the averaged spin of the electron than to its probability density. A particular setup for measuring the effect on spin dynamics, for example, in a laser field, might need accounting for other effects too \cite{ilderton2020loop}. Such a problem requires careful treatment and is out of the scope of this work and to be presented elsewhere. Moreover, further investigations are needed to solve Dyson-Schwinger equation for longer time beyond of the approximations made.

\begin{acknowledgments} 

We are grateful to E.~S.~Sozinov for the valuable discussions. This work was supported by the Russian Science Foundation (Grant No. 20-12-00077, calculation of the expectation value of the Dirac spin operator). A.A.M. was supported by the Russian Foundation for Basic Research (Grant No. 19-32-60084) and by Sorbonne Universit\'e in the framework of the Initiative Physique des Infinis (IDEX SUPER). A.M.F. was supported by the MEPhI Program Priority 2030. 
	
\end{acknowledgments}

\appendix

\section{Spin operator for the Dyson-Schwinger equation}
\label{appendix}

Let us come back to the construction~Eq.~\eqref{fp} of a spinor $\psi_\sigma (p)$ with positive energy and definite spin projection along the direction of $n$ with account for radiative corrections. Such a spinor should obey
\begin{eqnarray}
	\gamma^{5}\slashed{n} \psi_\sigma (p) & = & \sigma \psi_\sigma (p),
	\label{a1} 
\end{eqnarray}
which is equivalent to
\begin{eqnarray}
	\mathcal{L}_{-\sigma}  \psi_{\sigma} (p) & = & 0.
	\label{a3}
\end{eqnarray}
According to Eq.~\eqref{Ln} we have $\mathcal{L}_{-\sigma} \mathcal{L}_{\sigma} = 0$ for any $n$ such that $n^2=1$. However, in order to satisfy Eq.~\eqref{a3} in addition the operators  $\mathcal{L}_{-\sigma}(n)$ and $\overline{\mathcal{D}}$ should commute. Without account for radiative corrections this trivially holds for any $n$ such that $p\cdot n=0$. This just means that $\textbf{n}$ can be directed arbitrarily in a proper frame.

When radiative corrections are accounted for, this condition is more restrictive. Let us check that in a constant crossed field $\mathcal{L}_{-\sigma}(n)$ and $\overline{\mathcal{D}}$ [see Eq.~\eqref{eq:barD} with the coefficients defined as in Eqs.~\eqref{eq:defV}--\eqref{eq:defA}] commute for $n = n_D$, where $n_D$ is given by Eq.~\eqref{nD}. The nontrivial part of the commutator consists of the following terms:
\begin{equation}
	\begin{split}
		&\mathcal{L}_{-\sigma} \overline{\mathcal{D}} - \overline{\mathcal{D}} \mathcal{L}_{-\sigma}   =  
		\\
		&\quad\quad= \sigma\gamma^{5} \frac{n \cdot \mathcal{V} }{2} -\sigma \frac{ \slashed{n} \slashed{\mathcal{A}} - \slashed{\mathcal{A}} \slashed{n} }{4} + \sigma \gamma^{5} \frac{\slashed{n} \slashed{\mathcal{T}} -\slashed{\mathcal{T}} \slashed{n} }{4},
		\label{a4}
	\end{split}
\end{equation}
where $\slashed{\mathcal{T}} = \sigma^{\mu\nu}\mathcal{T}_{\mu\nu} = \gamma^\mu \gamma^\nu \mathcal{T}_{\mu\nu}$. For $n=n_D = \mathcal{A}/(a m \chi)$ we immediately have
$\slashed{n} \slashed{\mathcal{A}} - \slashed{\mathcal{A}} \slashed{n} = 0$. 

In virtue of Eq.~\eqref{eq:defV} the first term in \eqref{a4} for $n=n_D$ looks as follows:
\begin{equation}
	n_D \cdot \mathcal{V}  =   \frac{v_1 e}{\chi m^3} F^*_{\mu \nu} p^\nu p^{\mu} + \frac{v_2 e^3}{\chi m^7} F^*_{\mu \sigma} F^{\mu\nu}F_{\nu\lambda}p^\lambda  p^\sigma \;\;
	\label{a9}
\end{equation}
Here the first term in RHS vanishes due to antisymmetry of the tensor $F^*_{\mu \nu}$, while the second term vanishes due to the identity
\begin{equation}
	F^*_{\mu \sigma} F^{\mu \nu}=\epsilon_{\mu\sigma\alpha\beta}k^\alpha {A^\beta}'(\varphi)\left(k^\mu {A^\nu}'(\varphi)-k^\nu {A^\mu}'(\varphi)\right)=0,
	\label{A10}
\end{equation}
which holds for any plane wave in virtue of Eq.~\eqref{F-A} and the antisymmetry of the Levi-Civita symbol. 

Finally, applying the properties of the $\gamma$-matrices the last term in Eq.~\eqref{a4} simplifies to
\begin{eqnarray}
 \slashed{n}_D \slashed{\mathcal{T}} -\slashed{\mathcal{T}} \slashed{n}_D & \propto &  F^*_{\mu \nu} p^\nu F_{\alpha \beta} \left( \gamma^\mu \gamma^\alpha \gamma^\beta - \gamma^\alpha \gamma^\beta \gamma^\mu \right) \;\;
	\nonumber
	\\
	& = &  2  p^\nu \left( \gamma^\beta F^{* \mu \nu} F_{\mu \beta} -  \gamma^\alpha F^{* \mu \nu} F_{\alpha \mu } \right),
\end{eqnarray}
hence also vanishes in virtue of Eq.~\eqref{A10}.

Therefore, $\mathcal{L}_{-\sigma} (n_D)$ and $\overline{\mathcal{D}}$ indeed commute and hence Eqs.~\eqref{a3} and \eqref{a1} are fulfilled. As mentioned in the main text, $n_D$ is singled out by that in a rest frame its spatial component $\mathbf{n}$ is directed along the magnetic field.


%

\end{document}